\documentclass[graybox,envcountchap,openany]{svmono}

\usepackage{type1cm}         

\usepackage{makeidx}         
\usepackage{graphicx}        
\usepackage{multicol}        
\usepackage[bottom]{footmisc}
\usepackage{dcolumn}
\usepackage[flushleft]{threeparttable}
\usepackage{bm}
\usepackage{hyperref}
\usepackage{siunitx}
\usepackage{float}
\usepackage{multirow}
\usepackage{amstext}
\usepackage{babel}
\usepackage{tablefootnote}
\usepackage{longtable}
\usepackage{subcaption}
\usepackage{tikz}
\usepackage{pgfplots}
\usepackage{tikz-3dplot}
\usetikzlibrary{external}
\usepackage{afterpage}

\usepackage{newtxtext}       %
\usepackage{newtxmath}       

\usepackage[sectionbib]{natbib}
\usepackage{chapterbib}

\makeindex             

\graphicspath{{./}{images/}}

\begin{document}

\author{Carlos A. Salas \\ Neelima G. Kelkar}
\title{Astrophysical $S$-factors in primordial nucleosynthesis}
\subtitle{A comprehensive review}
\maketitle

\frontmatter

%
%

\preface

The astrophysical $S$-factor plays an important role in extrapolating the laboratory cross section data to lower energies of relevance in nucleosynthesis calculations. 
Be it empirical fits or theoretical modeling of the $S$-factors, experimental information on the cross sections over a range of energies is essential to provide reliable $S$-factors and thereby the nuclear reaction rates which enter the nucleosynthesis networks. 
In view of the scattered and sometimes diverse data on the same reaction in literature and the databases, we present a compilation and review of available data from 318 references involving 371 experiments on the leading and sub-leading reactions in big bang nucleosynthesis (BBN). Salient features such as bumps at certain energies and a steep rise in the $S$-factor at low energies due to screening effects are found to vary in different experiments. The latter depends on the variation of targets and sometimes temperature to extract information on the same nuclear reaction. For instance, 45 cases in which the S-factor at low-energy could present electron screening-effect showed deviations from background values with features which are unique to each experiment, target and temperature. 
The extensive compilation, discussions of the relevance of each reaction in BBN and the features of the $S$-factors is followed by a presentation of empirical fits to the $S$-factors in some selected reactions of interest. 

The book begins with an introduction to the astrophysical $S$-factor and some description of each of the different theoretical approaches used in the literature for its determination. A chapter on the basic concepts such as cross sections and nuclear reaction rates, essential for understanding the rest of the chapters, follows. Major part of the book focuses on the salient features of a widely compiled data on the astrophysical $S$-factor in the leading and sub-leading processes of big bang nucleosynthesis.
This is accompanied with an insight into the physics of each of these reactions and extensive literature.  We hope that it serves students in nuclear physics as well as researchers.

\vspace{\baselineskip}
\begin{flushright}\noindent
Bogot\'a, Colombia\hfill {\it Carlos Salas}\\
April 2026\hfill {\it Neelima Kelkar}\\
\end{flushright}

%
%

\extrachap{Acknowledgements}
The authors thank Marek Nowakowski for many useful discussions and suggestions. \\
\\
The authors also thank Kohji Takahashi, Helen Meyer, Roberta Spartà, Suqing Hou, Sergey B. Dubovichenko, Weijie Du, Sk M. Ali, Gerald M. Hale, Changbo Fu, Sophiya Taova, Dario Lattuada,  Naohiko Otsuka, Viktor Zerkin and Sonia Bacca  for providing reaction data and valuable information. \\
\\
This is a preprint of the following work: Carlos A. Salas and Neelima G. Kelkar, Astrophysical S-Factors in Primordial Nucleosynthesis, 2026, SpringerBriefs in Physics. It is the version of the author's manuscript prior to acceptance for publication and has not undergone editorial and/or peer review on behalf of the Publisher (where applicable). The final authenticated version is available online at: https://doi.org/10.1007/978-3-032-30466-7



\mainmatter
%
%
%
\chapter{Introduction}
\label{intro} 

Nuclear cross sections depend on the relative velocity of the projectile and target system which can vary over a range of values in an
astrophysical environment. The reaction rates which form an input to the network
of differential equations used to calculate the abundances of elements are given by integrals of the cross sections weighted by the probability distribution of the velocities. Accurate measurements or theoretical estimates of the low energy cross sections of nuclear reactions are thus crucial for precisely predicting the primordial abundances of light elements. However, cross sections at low energy rise sharply leading to difficulties with extrapolation of data to very low energies. Rewriting 
the cross sections of charged particle induced reactions, 
\begin{equation}\label{sigmandS}
    \sigma(E) = {1 \over E} \, \exp{(-2\pi\eta)} \, S(E)\,,
\end{equation}
removes the $1/E$ energy dependence and the {\it Gamow factor}, 
$\exp{(-2\pi\eta)}$, which is the leading term of the s-wave Coulomb barrier transmission coefficient for small energies compared to the Coulomb barrier height. The Sommerfeld parameter, $\eta$ is defined as $\eta = Z_1 Z_2 e^2 /(\hbar v)$ with $Z_1$ , $Z_2$ being the charges of the interacting nuclei. 
$S(E)$ is the ``astrophysical $S$-factor" which clearly varies far less strongly with energy as compared to the cross sections. 
The neutron induced cross section (for s-wave neutrons) is proportional to the neutron velocity, 
$\sigma \sim 1/v \sim 1/\sqrt{E}$ and one can write, $\sigma v \sim $ constant. In case the product $\sigma v$ is not constant one can introduce a function $R(E)$ and 
write $\sigma = R(E) /\sqrt{E}$. The $S$-factor in this case would be defined as 
$S(E) = \sigma(E) E = R(E) \sqrt{E}$.  
The $S$-factors will not be smoothly varying functions of $E$ for resonant reactions. The strong energy dependence in these cases is usually expressed in terms of a Breit-Wigner formula \cite{blattweisskopfbook}. 

Given the difficulty in measuring cross sections at very low energies, there exist several theoretical methods for estimating the astrophysical $S$-factor. Each method is more convenient depending on the nature of the reaction. For example, 
method accuracy can vary depending on the reaction type or the existence of 
resonant phenomena. The energy dependence of S(E) arises from the nuclear structure effects and is expected to be weak for nonresonant reactions. In view of this, in \cite{ueda2000} the authors obtained approximate analytic formulas for an effective $S$-factor, $S_{eff}$, by performing a Taylor series expansion of $S(E)$ about the threshold energy $E$=0 and about the Gamow energy, $E = E_0$. The method was extended in \cite{ueda2004} to obtain expressions for the thermonuclear reaction rates in charged particle collisions with a single narrow or broad resonance described by a Breit-Wigner form.

The cross sections at low energies are typically described with potential models via nucleus-nucleus potentials fitted to experimental data \cite{yakovlev2010,singh2019}. 
However, there do exist more sophisticated {\it ab initio} calculations where the 
nucleus is regarded as a many body system of interacting nucleons. One such calculation within Fermionic Molecular Dynamics (FMD) which uses a Gaussian
wave-packet basis to represent the many-body states was performed in \cite{neff2011}. 
The authors found a reasonably good agreement with the $^3$H($\alpha$, $\gamma$)$^7$Li $S$-factor data and very good agreement with the $^3$He($\alpha$, $\gamma$)$^7$Be data. A more recent {\it ab initio} calculation of the $^3$He($\alpha$, $\gamma$)$^7$Be with chiral two- and three-nucleon forces can be found in 
\cite{atkinson2025}. {\it Ab initio} no core shell model (NCSM) with realistic 
two- and three-nucleon interactions used to describe nuclear structure is extended 
to nuclear reactions in \cite{navratil2007}. Starting from {\it ab initio} wave functions of $^7$Be and $^8$B, the authors evaluate the overlap functions and overlap integrals with the objective of reproducing the 
$S$-factor for the $^7$Be(p, $\gamma$)$^8$B reaction. However, the $S$-factor also depends on the continuum wave function which, in the absence of an extension of the NCSM to the continuum is calculated using Wood-Saxon potential model. 

Clustering aspects of the structure of light nuclei can play an important role in 
the low energy domain of the nuclear reactions taking place shortly after the 
big bang. Some light nuclei of interest in the astrophysical context, such as 
$^6$Li, $^7$Li, $^7$Be, $^9$Be, $^{10}$B and $^{11}$B are known to exhibit a 
pronounced cluster structure. Ref. \cite{perrotta2023} 
presented an analysis of the $^6$Li(p, $^3$He)$\alpha$ transfer
reaction, based on first- and second-order distorted wave Born approximation (DWBA)  calculations. The authors also explored the impact of deformed components in the $^6$Li wave function in this reaction. It was found that the presence of 
clustered components in the wave function of $^6$Li led to a significant enhancement of the astrophysical $S$-factor whereas the static deformations of the ground state configuration played a negligible role at the low energies considered. 
The two-body potential cluster model of Ref. \cite{dubovichenko2018} presented a calculation of the $S$-factor above 10 keV for the radiative capture reaction $^3$He($^4$He, $\gamma$)$^7$Be to find good agreement with data. The authors also provide an analytical parametrization of the reaction rate for the same reaction. 

An interesting approach which can be used for resonant and non-resonant reactions is the $R$-matrix method which considers two regions: an internal one 
where the nuclear force is important, and the external one dominated by 
the Coulomb interaction between nuclei \cite{descouvement2005, descouvement2010}. The $R$-matrix at energy $E$, 
in a multichannel problem can be parametrized as a sum of $N$ pole terms characterized by energy $E_{\lambda}$ and width $\gamma_{\lambda}$ as \cite{descouvement2003}
\begin{equation}
    R_{ij} (E) = \sum_{\lambda = 1}^N \, {\tilde{\gamma}_{\lambda ,i} \tilde{\gamma}_{\lambda , j} \over E_{\lambda} - E} \,.
\end{equation}
Analysis of the $^3$He(n,p)$^3$H and $^7$Be(n,p)$^7$Li reactions within this approach is provided in \cite{descouvement2003}. Using direct and Trojan Horse method data, fits to the $S$-factor data of the $^2$H(d,p)$^3$H, $^2$H(d,n)$^3$He, $^7$Li(p,$\alpha$)$^4$He and $^3$He(d,p)$^4$He reactions were performed using the $R$-matrix public code AZURE \cite{azuma2010}.
Combining an hierarchical Bayesian model with an $R$-matrix model, the data on the $^7$Be(n,p)$^7$Li reaction which impacts the $^7$Be and $^7$Li abundances was analyzed in \cite{iliadis2020}. The authors provide $R$-matrix parameters. 
Parametrizations of the $S$-factors for several BBN reactions in terms of polynomials in energy $E$ and Breit-Wigner functions can be found in \cite{meissner2023}. 

The next chapter provides an introduction to all basic concepts essential to understand the technical details of the chapters which follow. Chapter 3 introduces the reader to primordial nucleosynthesis and the current problems in the calculation of the abundance of light elements. 
In Chapter 4, we present a compilation and review of available cross section data from 318 references involving 371 experiments on the leading and sub-leading reactions in big bang nucleosynthesis (BBN). 
To be specific, the $S$-factors for the leading and subleading processes in BBN are presented along with fits to the same. Salient features of these and peculiar features arising either due to the underlying physics or due to the particular experimental conditions are analyzed in detail. Experimental data was compiled by integrating standard databases such as EXFOR \cite{EXFOR_ref, EXFOR_web} and NACREII \cite{NACREII} with data collected by Goncharov (2018) \cite{Goncharov2018} on the $\mathrm{{}^2H}$(d,n)$\mathrm{{}^3He}$ and $\mathrm{{}^2H}$(d,p)$\mathrm{{}^3H}$ reactions, Serpico et al. (2004) \cite{serpico2004} on several BBN reactions, Spartá et al. (2020) \cite{Sparta2020} where the data was obtained by indirect methods, Du et al. (2022) \cite{du2022} on the $\mathrm{{}^1H}$(n,$\gamma$)$\mathrm{{}^2H}$ reaction, Soloyev (2023) \cite{Solovyev_2023} on the $^2$H ($\alpha$,$\gamma$) $^6$Li reaction, de Souza et al. (2019) \cite{deSouza2019},  Hupin et al. (2019) \cite{Hupin_Quaglioni_Navratil_2019}, and Nollet \& Burles (2000) \cite{Nollett_Burles_2000} on the $^3$H(d,n)$^4$He reaction, Varlachev et al. (2021) \cite{Varlachev_Dudkin_Nechaev_Penkov_Filipowicz_Philippov_Flusova_Chumakov_Shuvalov_2021}, Dubovichenko (2017) \cite{Dubovichenko_2017},  Dubovichenko et al. (2017) \cite{Dubovichenko_Dzhazairov-Kakhramanov_Afanasyeva_2017} and Dubovichenko \& Uzikov (2011) \cite{Dubovichenko_Uzikov_2011} on the $^3$H(p,$\gamma$)$^4$He reaction, Angulo et al. (2005) \cite{Angulo_Casarejos_Couder_Demaret_Leleux_Vanderbist_Coc_Kiener_Tatischeff_Davinson_2005} on the $^7$Be (d,p) $^8$Be reaction, Hou et al. (2015) \cite{Hou_He_Kubono_Chen_2015} on the $^7$Be (n, $\alpha$)$^{4}$He reaction,  de Souza et al. (2020) \cite{deSouza_Kiat_Coc_Iliadis_2020} and Adahchour \& Descouvemont (2003) \cite{Adahchour_Descouvemont_2003} on the $^7$Be(n,p)$^7$Li reaction, Sabourov et al. (2006) \cite{Sabourov_Ahmed_Blackston_Crowell_Howell_Perdue_Sabourov_Tonchev_Weller_Prior_Spraker_2006} and Hou et al. (2021) \cite{hou2021} on the $^7$Li (d,n) $^8$Be reaction, C. Spitaleri et al. (2019) \cite{Spitaleri_LaCognata_Lamia_Pizzone_Tumino_2019}, Taova et al. (2017) \cite{Taova_Selyankina_Generalov_Zherebtsov_Lipenkova_Tulina_2017} on the $^7$Li (p,$\alpha$)$^4$He reaction, and the relevant references therein. Later, the compilation was extended to data from recent experiments on the selected BBN reactions as of December 2025. 
Chapter 4 also gives insights into the different processes and their relevance in BBN. 
At the same time, the data and corresponding fits are displayed in the figures in this section. Some fits were adapted from the formulae used in the paper by Meissner et al. (Appendix A) \cite{meissner2023}.
The extensive compilation of data allowed us to note that the screening effect is influenced by the target and temperatures in a particular experiment. 
A detailed discussion of the available cases in literature is presented in Chapter 5. 

\bibliographystyle{unsrt}
\bibliography{bibliography}





\end{document}